# Using the Active Collimator and Shield Assembly of an EXIST-Type Mission as a Gamma-Ray Burst Spectrometer

A. Garson III[1], H. Krawczynski[1], J. Grindlay[2], G.J. Fishman[3], and C.A. Wilson[3]

[1] Washington University in St. Louis, 1 Brookings Drive, CB 1105, St. Louis, Mo, 63130, USA
[2] Harvard-Smithsonian Center for Astrophysics, 60 Garden Street, Cambridge, MA 02138, USA
[3] National Space Science & Technology Center, 320 Sparkman Dr., Huntsville, AL 35805 USA

**Abstract.** The *Energetic X-Ray Imaging Survey Telescope (EXIST)* is a mission design concept that uses coded masks seen by Cadmium Zinc Telluride (CZT) detectors to register hard X-rays in the energy region from 10 keV to 600 keV. A partially active or fully active anti-coincidence shield/collimator with a total area of between 15 $m^2$ and 35 $m^2$ will be used to define the field-of-view of the CZT detectors and to suppress the background of cosmic-ray-induced events. In this paper, we describe the use of a sodium activated cesium iodide shield/collimator to detect gamma-ray bursts (GRBs) and to measure their energy spectra in the energy range from 100 keV up to 10 MeV. We use the code GEANT 4 to simulate the interactions of photons and cosmic rays with the spacecraft and the instrument and the code DETECT2000 to simulate the optical properties of the scintillation detectors. The shield/collimator achieves a $\nu F_\nu$-sensitivity of $3 \times 10^{-9}$ erg cm$^{-2}$ s$^{-1}$ and $2 \times 10^{-8}$ erg cm$^{-2}$ s$^{-1}$ at 100 keV and 600 keV, respectively. The sensitivity is well matched to that of the coded mask telescope.The broad energy coverage of an EXIST-type mission with active shields will constrain the peak of the spectral energy distribution (SED) for a large number of GRBs. The measurement of the SED peak may be key for determining photometric GRB redshifts and for using GRBs as cosmological probes.

## 1. Introduction

Gamma-ray bursts (GRBs) remain one of the most active and most interesting fields of high energy astrophysics. "Long GRBs" (durations longer than 2 sec), for which afterglows have been detected, are believed to originate from "hypernovae", or collapsars. In this model, the core of a massive star progenitor collapses into a black hole that subsequently accretes a substantial fraction of the star's mass, giving rise to a highly relativistic collimated outflow (jet) with an initial bulk Lorentz factor on the order of 100-1000 (Woosley 1993, Pacynski 1998). Internal shocks due to an intermittent outflow and external shocks generated by the relativistic outflow sweeping up ambient material might generate the prompt and afterglow emission, respectively. The hypernova hypothesis is supported by the observation of a supernova-type optical spectrum associated with GRB030329 (Stanek et al. 2003) and a correlation of the GRB redshift distribution with the cosmic star-formation history, see e.g. (Lloyd-Ronning et al. 2003). "Short GRBs" (durations shorter than 2 sec) might originate from a different type of catastrophic event, such as neutron star mergers (Narayan et al. 1992). GRB observations make it possible to study the astrophysics of these powerful explosions. Given the enormous luminosities of GRBs, up to several $10^{51}$ ergs s$^{-1}$ sr$^{-1}$, future observations of GRBs with redshifts on the order of 10 might allow us to study star formation rates at high $z$, the formation of stellar mass black holes from the collapse of population III stars, and the interstellar medium environment of cosmologically located star forming regions. Recent reviews of GRB observations and models can be found in Djorgovski et al. (2003), Waxman (2003), Frontera et al. (2004), and Meszaros (2002).

In this paper, we discuss a new GRB detector which builds on a long legacy of GRB missions. In 1991, the *Burst and Transient Source Experiment (BATSE)* on board the *Compton Gamma-Ray Observatory* (CGRO, 1991-2000) was launched. The experiment consisted of eight detector modules, each carrying one thin and one thick NaI scintillation detector. The thin (1.3 cm), 51 cm diameter detectors of each module were used to detect GRBs, determine their location in the sky, and detect lightcurves; the thick (7.6 cm), 12 cm diameter detectors were used to determine the GRB energy spectra over the broad energy range from 5 keV to 2 MeV. The *BeppoSAX* (Frontera 2004) and *HETE* (Lamb et al. 2004) missions located GRBs with arcmin accuracy and led to the detection of afterglows by ground-based and space-borne observatories.

In late 2004, the GRB observatory *Swift* was launched. Its Burst Alert Telescope (BAT) uses CZT detectors and the "coded mask" approach to detect GRBs in the energy range from 15 to 150 keV and to obtain GRB localizations of 4 arcmin accuracy.

After slewing the spacecraft rapidly toward the GRB direction, the X-ray Telescope (XRT) is used to obtain 5 arcsec localizations. In the coded mask approach, detectors see the sky through a patterned shadow mask. The detectors image the sum of all shadow-images cast by the X-ray sources in the field-of-view. A deconvolution algorithm is used to derive the X-ray surface brightness distribution from the detected image. Swift has made already substantial contributions to GRB science, including the detection of a GRB at z=6.29 (Haislip et al. 2006) and the afterglow and counterpart study of short and long GRBs (Gehrels et al. 2004).

The 2003 roadmap of the NASA theme "Structure and Evolution of the Universe", the Beyond Einstein program, recommends a Black Hole Finder Probe (BHFP) to conduct an all-sky survey for black holes. The *EXIST* telescope (Grindlay et al. 2005) is one possible realization of the BHFP. The present design combines a high-energy (HE) detector (15-600 keV) with a low-energy (LE) detector (5-30 keV), both being coded mask, wide field of view telescopes. Scanning the entire sky each 95 min orbit, *EXIST* will survey the sky for the emission from obscured and un-obscured active galactic nuclei (AGNs), GRBs, and galactic black hole systems. The LE telescope complements the HE telescope by improving the source localization accuracy from 1 arcmin to 10 arcsec. *EXIST's* sensitivity for GRBs will be a factor of 10 higher than that of Swift, and a GRB rate of between two and three per day with localizations of better than 50 arcsec is anticipated. Importantly, *EXIST* will have extremely high sensitivity at energies above 100 keV, enabling the sensitive detection of the more enigmatic short GRBs which exhibit harder energy spectra than long GRBs (Dezalay et al. 1992).

In this paper, we discuss the possibility if using a fully active CsI(Na) collimator/shield of an *EXIST*-type mission, to detect gamma-ray bursts and to measure their energy spectra. While the instrument design is still in a preliminary stage, total detector area is expected to be between 15 $m^2$ and 35 $m^2$. This collimator/shield can thus achieve an excellent sensitivity and can complement the CZT detectors at energies above $\sim 300$ keV.

Several satellite borne experiments have used their anti-coincidence shields as GRB detectors. It is believed that HEAO A-4 was the first gamma-ray astronomy experiment that used a large active side shield as a GRB detector. The SIGMA telescope on-board the GRANAT satellite had an anti-coincidence shield of 8 CsI(Na) blocks, with each 4 cm thick block having a detection area of 0.15 $m^2$ (Paul et al. 1990). In the years from 1990 to 1992, the active shield recorded high quality 200 keV-15 MeV energy spectra for 25 GRBs (Pelaez et al. 1994). The OSSE instrument on the CGRO spacecraft also had a large shield composed of NaI scintillator that could be used to monitor GRBs. It had a time resolution of 64 ms. The GRB Monitor (GRBM) on-board the satellite BeppoSAX (Costa et al. 1998). The primary use of its four CsI(Na) lateral shields (each 0.114 $m^2$ area, 1 cm thick) is to shield the Phoswich Detector System (PDS). In addition, the shield has been used to detect GRBs and to measure their light curves and energy spectra. Each lateral shield is read out by two Photo-Multiplier Tube (PMTs). The GRBM has been used to measure energy spectra in the energy region from 40 to 700 keV. Its peak flux sensitivity is $10^{-8}$ erg cm$^{-2}$ s$^{-1}$ yielding 10 GRB triggers per month (Feroci et al. 1997, Amati 2002). Although only a secondary experiment, the GRBM has been extremely successful.

In more recent times, the large segmented BGO shield for the SPI instrument on the INTEGRAL spacecraft has been very successful in providing sensitive GBB observations. The geometry, capabilities, sensitivity, and observations from the SPI shield detector is described by Ryde, et al. (2003). The instrument was launched in October 2002 and continues to be in operation. Compared to the earlier telescopes, the partially or fully active detector/collimator/shield assembly of an *EXIST*-type mission will have a very different geometry (different shadowing, different projected slab thicknesses) and a much larger effective area than the shield detectors mentioned above.

This paper is organized as follows. In Sect. 2 we describe the telescope layout and the detector simulations. In Sect. 3 we calculate the effective area and energy resolution of the active shield and we estimate the different sources of background, e.g. the diffuse X-ray background, albedo emission, trapped protons and electrons, and cosmic-ray protons and electrons. In addition, we discuss the GRB sensitivity of the collimator/shield and investigate the accuracy with which GRB energy spectra can be reconstructed. We will conclude with a summary and a discussion of the results in Sect. 4.

## 2. Telescope Design and Simulation Details

NASA has selected *EXIST* for a 2-year concept study, along with the *Coded Aperture Survey Telescope for Energetic Radiation (CASTER)*, a competing design which employs the new scintillators LaBr$_3$ or LaCl$_3$ as primary detectors (Cherry et al. 2004). As the *EXIST* mission is foreseen to be launched in the year 2020, but with large uncertainties, the design is still evolving. However, a smaller version of *EXIST* may be launched earlier. We limit our analysis to an approximation of the full-size *EXIST* geometry as of summer 2004. The main emphasis of our study is to determine the energy range over which the collimator/shield will give useful spectral constraints. Another effect of great interest is the relative importance of "side" and "bottom" slabs. We will explain those terms further below.

We limit the study to the simulation of the HE telescope (see Fig. 1). The design divides the instrument into three "telescopes", each with a 60° by 75° field of view which combine to give the entire instrument a 180° × 75° total field of view. Each of the three telescopes is composed of nine "sub-telescopes". In our simplified treatment, each sub-telescope is basically an open rectangular box (see Figure 2). While we assume a design based on right angles, the actual *EXIST* design joins slabs at angles slightly larger than 90°, resulting in a somewhat better photon detection efficiency. In the following we refer to the open sides of the boxes as "top" and the closed side of the box as "bottom".

The 7 mm thick tungsten coded mask is positioned ∼1.5 m from the bottom of each telescope. An assembly of $25 \times 29$ CZT detectors (each W × D × H of $2.0 \times 2.0 \times 0.5$ cm$^3$) covers an area of $50.0 \times 58.0$ cm$^2$ at the bottom of the sub-telescope, and views the coded mask. The CZT detectors are shielded by a 2 cm thick and $50.0 \times 58.0$ cm$^2$ large bottom detector and by four 1 cm thick side slabs. Two opposing side slabs have a size of $45.0 \times 60.0$ cm$^2$; the other two opposing slabs are $45.0 \times 52.0$ cm$^2$ large (both W × H). If not stated otherwise, we assume that both slabs are fully active and are made of CsI(Na). We chose CsI(Na) owing to its durability, high density, and high light output. Sodium activated CsI does not show the "slow" ($\mu$sec) scintillation phenomenon that plagues Tl activated CsI in space-borne applications. Due to the hygroscopic nature of CsI(Na), we assume each slab is sealed in an aluminum casing with quartz windows for optical coupling. We do not foresee this restricting the instrument design or cost. The bottom slab is subdivided into four optically independent CsI slabs each read out by one 3" Photo Multiplier Tube (PMT). Allowing for coupling inefficiencies, we use a conservative quantum efficiency of 15% and an excess noise factor of 1.2. We assume monolithic side slabs, each read out by two waveshifter bars located at the top and the bottom of the slabs. The waveshifters are modeled similar to the Saint-Gobain[1] waveshifter BC-482A. We use an efficiency of 80% for waveshifting events. The mean free path inside the waveshifter for scintillation and waveshifted photons are 2 mm and 1 m, respectively. The waveshifter bars are held a short distance from the scintillator slabs to increase internal reflection inside the bars. Each waveshifter is read out by one hybrid PMT located at one side; we assume that the other side is sealed by a thin metal layer evaporated onto the end of the bar. We assume that the scintillator slabs and waveshifter bars are polished on all 6 sides, except the metal-coated end of the waveshifter bar, and are wrapped in white painted Aluminum foil. The optical coupling is done through quartz-glass windows.

The simulation is divided into two parts. Particle interactions are simulated with the GEANT 4 package (Agostinelli 2003) developed at CERN for high energy physics experiments. We use the Detect2000 code (Knoll et al. 1988, Tsang et al. 1995) to simulate the photon transport inside the scintillators and to the detectors. The GEANT 4 simulations include the low-energy electromagnetic processes package GLECS (Kippen 2004) for the simulations of low-energy interactions. GLECS accounts for the binding energy of electrons in Compton and Raleigh scatterings. The simulations incorporate the shield, the CZT detectors, and the coded mask for all 3 telescopes. We assume that the mask is made of 7 mm thick, $1 \times 1$ cm$^2$ large tungsten pieces covering 50% of the mask area. The true tungsten mask would have smaller openings. However, simulating the full mask becomes very time-consuming. For each GEANT 4 energy deposition into a scintillator slab, scintillation photons are generated taking into account the non-linear light yield characteristics of CsI(Na) described by Mengesha et al. (1998).

The Detect2000 code simulates photon transport inside a wide variety of optical materials taking into account the index of refraction, photon scattering length, photon attenuation length, surface finish, and diffuse surface reflectivity. Photons are isotropically generated within the material and tracked on an individual basis. The program logic calculates the path length to the next intersection of a surface. Random sampling then determines if the photon is absorbed, scattered, or wavelength shifted over this path. If none of these occur, the optical properties of the next suface determine whether the photon is reflected, refracted, detected, or absorbed. We use the code to transport the photons through the scintillator slabs, the waveshifter bars, and the photon detector entrance windows. As the last step in the simulations, photons reaching the detectors are converted into PMT signals.

Figure 3 shows four different slab/photodetector configurations that we use in Detect2000 simulations. In Figure 4 we plot the fraction of detected photons for these configurations. In the case of the bottom slab read out by four large photo-detectors (configuration A), averaging over the entire volume of the bottom slab gives a mean fraction of 3% of generated scintillation photons arriving at the large PMTs. In the case of the side slab read out with two waveshifter bars (configuration B) and two detectors, a mean fraction of 5% reaches the photodetectors. If the side slab is only read out by one waveshifter bar and one photo tube (configuration C), the fraction goes down to 2%. Configuration D replaces the two waveshifter bar/photodetector combinations of configuration B with 6 photo-detectors directly coupled to the scintillator slab. We find that configurations B gives much better performance than configuration D. While configuration B requires 3 times less power and readout channels than configuration D, the mass of both configurations is similar. In the following we assume that the side slabs are read out with configuration B.

Table 1 summarizes the characteristics of the slabs, waveshifter bars, and PMTs used as input to the simulation. We cross-checked the input parameters by verifying that the simulations correctly predicted the spectroscopic performance of the fully active shield of the BeppoSAX PDS instrument (Lorenzo 1998).

## 3. Results

*Effective Area and Energy Resolution*

In the following, we discuss the sensitivity of the shield/collimator assembly for GRBs. Rather than exploring the full regime of GRB parameters, we limit our study to GRBs of 30 sec duration, incident from a direction $40°$ off center of the field-of-view of Telescope #1 (see Figure 1). In Figure 5, the effective area for detecting 80%, 90%, and 95% of the primary photons' energy in any one slab is shown as function of the primary photon's energy. At about 250 keV, the Compton effect starts to dominate over the photoelectric effect, and the effective area decreases rapidly with increasing energy. Above ~1 MeV, the detection area shows a shoulder owing to pair production processes. At the highest energies, the detection area decreases due to the limited stopping power of the slabs.

---

[1] Saint-Gobain Crystals 12345 Kinsman Road, Newbury, OH 44065

Figure 6 shows the PMT signal amplitude as function of the energy of the primary photon. One can recognize 2 different regimes: at energies below 100 keV the energy resolution is limited by the statistical fluctuations of the number of scintillation photons and improves with increasing energy. Above 1 MeV, the energy resolution deteriorates with increasing energy as an increasing fraction of the photon's energy leaves the slabs. The $\sigma_{\ln(\Delta E/E)}$ is 15%, 14%, 13%, and 28% at 200 keV, 400 keV, 600 keV, and 800 keV, respectively. These values were calculated without optimization, treating all slabs on the same footing, and not correcting for orientation.

### Background Rates

The sensitivity of both the CZT detector and the shield/collimator assembly is limited by background fluctuations, rather than by photon count statistics. The dominant background sources are the diffuse extragalactic X-ray background, albedo photons from Earth's atmosphere, trapped protons and electrons, cosmic ray protons and electrons, prompt neutron reactions, and delayed activation. For the shield/collimator assembly, the background is dominated by diffuse X-rays and atmospheric X-rays; a simple estimate using the methods of Matteson et al. (1977) shows that prompt neutron reactions and activation can be neglected. In case of the "shielded" CZT detectors, external backgrounds are reduced, and prompt neutron reactions and activation become relatively more important. As this paper is only concerned with the performance of the shield/collimator and not with the performance of the CZT detectors, we will neglect prompt neutron reactions and activation. Using similar assumptions, Shaw et al. (2003) described the background of the BATSE experiment satisfactory.

We will assume a 550 km orbit at 7° inclination. The orbit is realistic for a spacecraft weight of 8500 kg and a launch vehicle of the class Delta IV(4050H) (Grindlay, 2004, private communication). The particles were generated with a random position and trajectory on a sphere of 500 m centered on the middle telescope. All detections triggered the filling of histograms according to total deposited energy and total PMT counts for each slab encountered per incident particle. The histograms were weighted according to the particle's differential flux for a 30 second duration. The histograms were then summed to produce a differential total background for bottom slabs and for side slabs.

The simulations use the diffuse extragalactic X-ray background as measured by the HEAO-1 satellite (Kinzer et al. 1997). The differential flux, $dN_d/dE$, is given by

$$\frac{dN_d}{dE} = 2.62 \times 10^{-3} \times \left(\frac{E}{100 \text{ keV}}\right)^{-2.75} \text{ cm}^{-2} \text{ s}^{-1} \text{ sr}^{-1} \text{ keV}^{-1}. \tag{1}$$

At 100 keV, 1 MeV, and 10 MeV the differential flux is $2.2 \times 10^{-3}$, $4.5 \times 10^{-6}$, and $1.5 \times 10^{-8}$ cm$^{-2}$ s$^{-1}$ sr$^{-1}$ keV$^{-1}$, respectively.

Cosmic ray interactions with Earth's atmosphere cause a flux of atmospheric X-rays and gamma-rays, commonly referred to as "albedo". The simulations use the simplified spectrum of Longo et al.(2002), based on balloon flight data and SAS-2 data. This spectrum, representing the average of the emission over the solid angle subtended by the earth's surface at 550 km, is given by:

$$\frac{dN_a}{dE} = 8 \times 10^{-5} \times \left(\frac{E}{1 \text{MeV}}\right)^{-1.4} \text{ cm}^{-2} \text{ s}^{-1} \text{ sr}^{-1} \text{ keV}^{-1} \tag{2}$$

for $E \leq 10$ MeV, and

$$\frac{dN_a}{dE} = 5 \times 10^{-4} \times \left(\frac{E}{1 \text{ MeV}}\right)^{-2.2} \text{ cm}^{-2} \text{ s}^{-1} \text{ sr}^{-1} \text{ keV}^{-1} \tag{3}$$

for $E > 10$ MeV, where $dN_a/dE$ is the number of atmospheric photons at energy $E$.

We used ESA's SPace ENVironment Information System (SPENVIS)[2], to compute the time averaged cosmic ray proton flux and to determine the trapped electron and proton fluxes . We assume in the following that the science instruments are not used during the passage through the South Atlantic Anomaly (SAA). For the orbital parameters considered here, the trapped electron and proton fluxes are negligible during 92% of the orbit.

We used the cosmic electron flux from Gehrels (1985):

$$\frac{dN_e}{dE} = 2.27 \times 10^{-4} \times \left(\frac{E}{1 \text{ GeV}}\right)^{-3.086} \text{ cm}^{-2} \text{ s}^{-1} \text{ sr}^{-1} \text{ keV}^{-1} \tag{4}$$

with a low-energy cutoff at 2.5 GeV.

---

[2] http://www.spenvis.oma.be

The differential and integral background rates per sub-telescope (consisting of one bottom slab and 4 side slabs) outside the SAA are given in Figure 7. These rates are computed by averaging the background rates from all side slabs or all bottom slabs. These average rates are used to compute the rate of one sub-telescope. The background is dominated by the extragalactic X-ray background, Albedo radiation, and cosmic ray protons with the first dominating below 100 keV, the second dominating between 100 keV and 5 MeV, and the latter above 5 MeV. Cosmic ray electrons contribute noticeably only at energies above 30 MeV. The integral background trigger rate per sub-telescope is $2 \times 10^4$ Hz for a shield trigger threshold of 100 keV. Assuming that each background hit vetoes the CZT detectors of a sub-telescope for $5\mu$sec, the background rate would correspond to a dead time of 10%. It would be possible to use the shield as GRB detector with a threshold of 100 keV, and to use it as anti-coincidence detector with a higher "veto-threshold". For a veto-threshold of 1 MeV, the background rate would be $4 \times 10^3$ Hz and the dead time would be 2%. Alternatively, a partly active, partly passive shield combination could be used to reduce the dead time.

*Sensitivity*

We computed the GRB sensitivity of the shield/collimator assembly by simulating emission with a $dN/dE \propto E^{-2}$ energy spectrum over factor-of-2 wide energy intervals (e.g., 100-200 keV, 200-400 keV, ...). We adjusted the flux normalization until the GRB counts exceeded the background counts with a statistical significance of 5 standard deviations.

Counts were considered only if they fell into pre-determined PMT amplitude ranges, optimized for detecting photons in the energy interval under consideration. The results are shown in Figure 8. The side slabs are substantially more sensitive than the bottom slabs mainly because their total area is larger. Other effects are that the mask shields the bottom slabs more than the side slabs, and the CZT shields the bottom slabs at low energies.

The sensitivity of the shield matches the sensitivity of the CZT detectors (bold dashed line) in the region of overlap from $\sim 125$ keV to $\sim 250$ keV. The result shows that the two detectors complement each other ideally: the CZT detector will measure the low-energy end of the spectrum and will determine the GRB location with arcmin-accuracy. The shield/collimator will measure the energy spectrum from $\sim 200$ keV up to $\sim 30$ MeV for the very strongest bursts.

*Spectral Fits*

In this paragraph, we discuss the accuracy with which the shield/collimator assembly can measure the Spectral Energy Distributions (SEDs) of GRBs. We assume that the GRB energy spectrum can be described with the Band model (Band et al. 1993):

$$\frac{dN}{dE} = N_0 \times \left(\frac{E}{1\,\text{keV}}\right)^{\alpha} \exp\left(\frac{-\text{E}\,(2+\alpha)}{\text{E}_{\text{p}}}\right) \text{ for E } < \frac{(\alpha-\beta)\,\text{E}_{\text{p}}}{(2+\alpha)} \tag{5}$$

and

$$\frac{dN}{dE} = N_0 \times \left(\frac{(\alpha-\beta)\,E_{\text{p}}}{(1\,\text{keV}\,(2+\alpha))}\right)^{\alpha-\beta} \exp\left(\beta-\alpha\right) \left(\frac{\text{E}}{1\,\text{keV}}\right)^{\beta} \text{ for E } \geq \frac{(\alpha-\beta)\,\text{E}_{\text{p}}}{(2+\alpha)} \tag{6}$$

where E is the photon energy, $E_{\text{p}}$ is the the energy at which the SED peaks, $N_0$ is a normalization constant in photons cm$^{-2}$ s$^{-1}$ keV$^{-1}$ , and $\alpha$ and $\beta$ are the low and high-energy spectral indices, respectively. In the following, we do not quote $N_0$, but rather $N_{\text{P}}$, the $\nu$ $F_{\nu}$-flux at the peak of the SED (ergs cm$^{-2}$ s$^{-1}$). We study the spectroscopic performance of the shield/collimator for twelve combinations of $E_{\text{p}}$ and $N_{\text{P}}$ (see Table 2), assuming for all models typical spectral index values of $\alpha$ = -1 and $\beta$ = -2.5 (Preece et al. 1998). For each parameter combination we "simulate" 100 GRBs of 30 sec duration, incident, as in the previous paragraph, from a direction $40°$ off the center of the field-of-view of Telescope 1.

For each simulated GRB, we determine best-fit model parameters ($E_{\text{p}}$, $N_{\text{P}}$, $\alpha$, and $\beta$) with a forward folding approach. The fit uses two PMT pulse height histograms (PHHs). The first is filled with all the PMT pulse heights from the side slabs, and the second with all the PMT pulse heights from the bottom slabs. An individual GRB is simulated by filling the two histograms with Monte Carlo events, giving them a weight according to the assumed Band model. Noise is added, taking into account the signal and background count-statistics. The actual fit is performed by a search in the parameter space, comparing "template" pulse height histograms (filled by properly weighting the Monte Carlo events; this time no noise is added), with the pulse height histograms of the simulated burst for each parameter combination. The comparison uses the $\chi^2$-statistics. The search yields the parameter combinations which minimizes the $\chi^2$-value.

Figure 9 shows the histograms corresponding to the "simulated data" for all side slabs (upper histogram) and all bottom slabs (lower histogram) for a specific parameter combination $E_{\text{P}}$ = 450 keV, $N_{\text{P}}$ = $10^{-7}$ergs cm$^{-2}$ s$^{-1}$, $\alpha$ = $-1.0$, and $\beta$ = $-2.5$. The error bars represent the statistical errors, the square root of background counts + GRB counts for each bin . In the specific example, the best-fit model parameters were as follows: $E_{\text{P}}$ = $452 \pm 4$ keV, $N_{\text{P}}$ = $9.98 \pm .02 \times 10^{-8}$ergs cm$^{-2}$ s$^{-1}$, $\alpha$ = $1.02 \pm .01$, and $\beta$ = $-2.51 \pm -.006$ (errors are on a 90% confidence level).

Based on the best-fit model parameters for 100 simulated GRBs, the accuracy of determining the model parameters is computed. The results are summarized in Table 2 . A graphical representation is given in Figure 10. The results show that the model parameters can be determined with good statistical accuracy for a wide range of peak energies $E_p$ and flux levels $N_p$. Using $E_P = 450$ keV, $N_P = 10^{-7}$ergs cm$^{-2}$ s$^{-1}$, $\alpha = -1$, and $\beta = -2.5$, we also computed the distribution of resulting fit values for the spectral indices $\alpha$ and $\beta$ for 100 bursts. We find that 90% of the reconstructed $\alpha$-values lie within the interval -1.039 to -0.938, and 90% of the reconstructed $\beta$-values lie within the interval -2.562 to -2.418, showing that the spectral indices can also be determined with good statistical accuracy. Table 3 displays the results.

Some GRB spectra also exhibit characteristics of a blackbody spectrum (Ryde 2004). In addition to using the Band model, we also produced spectra with the Planck function, given by

$$\frac{dN}{dE} = N' \times \frac{(E/\mathrm{keV})^2}{\exp[E/kT] - 1} \tag{7}$$

where $k$ is the Boltzman constant ($8.617 \times 10^{-8}$ keV K$^{-1}$) , $T$ is the temperature in Kelvin, $E$ is the photon energy, and $N'$ is the flux amplitude normalization in photons cm$^{-2}$ s$^{-1}$ keV$^{-1}$. Letting $kT = 100$ keV, we normalized $N'$ so $N_P$, the $\nu$ $F_\nu$-flux at the peak of the SED, was $1.0 \times 10^{-7}$ergs cm$^{-2}$ s$^{-1}$ . Using the same method as above, we simulated 100 bursts of 30 second duration and fit $kT$ and $N_P$. We find that 90% of the reconstructed $kT$-values lie within the interval 98.7 to 102.4 keV and 90% of the reconstructed $N_P$-values lie within the interval 9.8 $\times 10^{-6}$ to 1.02 $\times 10^{-7}$ergs cm$^{-2}$ s$^{-1}$. The results show that the Planck model parameters can be determined with good statistical accuracy. Table 3 shows the distribution of the resulting fit parameters.

## 4. Discussion

The *EXIST* mission will detect GRBs with unprecedented sensitivity. While the Si and CZT detectors detect GRBs over the energy range from 5 keV to ~300 keV, the simulations described in this paper show that partially active shield/collimator CsI(Na) detectors can contribute spectroscopic information in the energy range from ~ 300 keV up to ~ 10 MeV. We find the sensitivity of the shield is well matched to the sensitivity of the CZT detectors. A fully active shield of an *EXIST* -type mission would have an effective area of ~ 55,000 cm$^2$, ~ 40,000 cm$^2$, and ~ 20,000 cm$^2$ at 200 keV, 400 keV, and 1 MeV, respectively. This is a greater area than BATSE's 8 Large Area Detectors (LADs), each having an effective area of ~ 1400 cm$^2$, ~ 900 cm$^2$, and ~ 550 cm$^2$, at the same respective energies (Pendleton et al. 1999).

The contributions of a shield/collimator assembly are important for trade off considerations concerning the science instrument design. A shield/collimator with good spectroscopic performance above 300 keV might reduce the necessity for a thick CZT detector/mask combination. Thicker CsI(Na) slabs will result in a better spectroscopic performance at higher energies and in reduced background leakage through the shield/collimator. However, thick shields are massive and expensive and at some point, the extra material will result in a background increase caused by activation of the CsI(Na) slabs themselves.

One design option that is currently considered by the *EXIST* team, is to make the shield partially active and partially passive by making the bottom slab and the lower half the side slabs from active CsI, and the upper half of the side slabs from passive material. Active shielding is relatively more important close to the detectors, as only the active shielding can veto prompt and delayed background events produced in the CZT detectors. From Fig. 8, one can see that the GRB sensitivity of the bottom and side slabs differs by an order of magnitude at 250 keV and a factor of two at high energies. Reducing the active area of the side slabs by a factor of two could thus change the overall sensitivity of the shield/collimator by approximately a factor $\sqrt{2}$.

The shield/collimator assembly will be instrumental for addressing a wide range of GRB science topics. The SEDs of BATSE GRBs peak in the region between 100 keV and 1 MeV (Mallozzi et al. 1995, Brainerd 1998). Spectroscopic coverage above Swift's 150 keV high energy limit is thus essential for determining the total energy flux and apparent luminosity of the prompt GRB emission.

At the time of writing this paper, redshifts have been determined for ~ 45 GRBs. Based on various sub-samples of these GRBs, several groups have found at least tentative evidence for correlations of several key GRB parameters. Norris et al. (2000, 2002) found evidence for a correlation between a lag between the low and high energy gamma-ray emission and the apparent luminosity. Fenimore & Ramirez-Ruiz (2000) and Reichart et al. (2001) reported a correlation between the flux variability amplitude and the apparent isotropic luminosity. More recently, several groups found evidence for a correlation of the energy at which the spectral energy distribution peaks $E_P$ (after $k$ correction), and:

- the apparent isotropic luminosity (Amati et al. 2002; Lloyd & Ramirez-Ruiz 2002),
- the apparent isotropic peak luminosity (Yonetoku et al. 2004), and
- the luminosity in the frame of the relativistically moving jet plasma (Ghirlanda et al. 2004).

While the correlations between $E_P$ and the apparent isotropic luminosity and apparent isotropic peak luminosity seem to be inconsistent with a considerable fraction of GRBs detected by BATSE (Nakar & Piran, 2004, Band & Preece, 2005), the last correlation is consistant with the data. A fully or partly active shield on an *EXIST*-type mission would allow us to determine $E_p$ for a large number of GRBs, and thus to test these correlations and similar correlations in great detail. Extensive studies of these correlations, and the deviations from these correlations as function of other GRB parameters, will be an important aim of a next-generation GRB mission. The shield will this play an important role in determining photometric redshifts of GRBs.

## *Acknowledgements*

We thank J. Matteson for very stimulating discussions. AG thanks the GEANT team and Saint-Gobain for technical support. This work has been supported in part by the McDonnell Center for Space Sciences at Washington University.

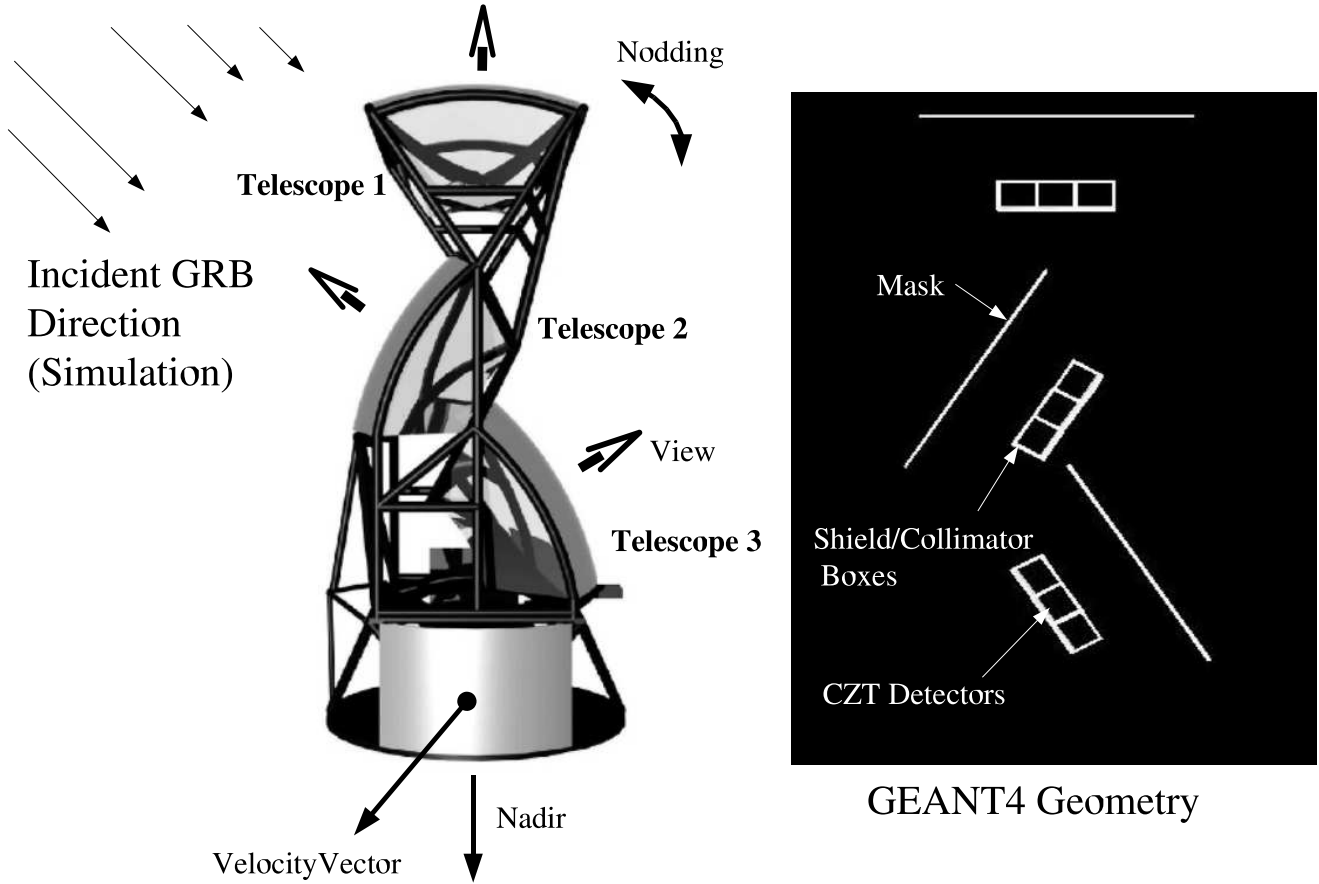

**Fig. 1.** Side view of the overall design for *EXIST* as of Summer 2004 (left side). Three telescopes, each with $60° \times 75°$ field of view, use the coded mask approach to image X-ray sources with CZT detectors. The satellite will allow for nodding about its direction of motion. The nodding will be used to correct for non-uniformities of single CZT detector units. The nadir direction points to Earth, while the velocity vector points in the direction of the orbit. The figure on the right shows the geometry used for GEANT4 simulations. The simulations included the coded masks (seen edge-on), the shield/collimator assemblies (white boxes), and the CZT detectors (not visible). The incoming photons at the upper left indicate the incident direction of simulated GRBs.

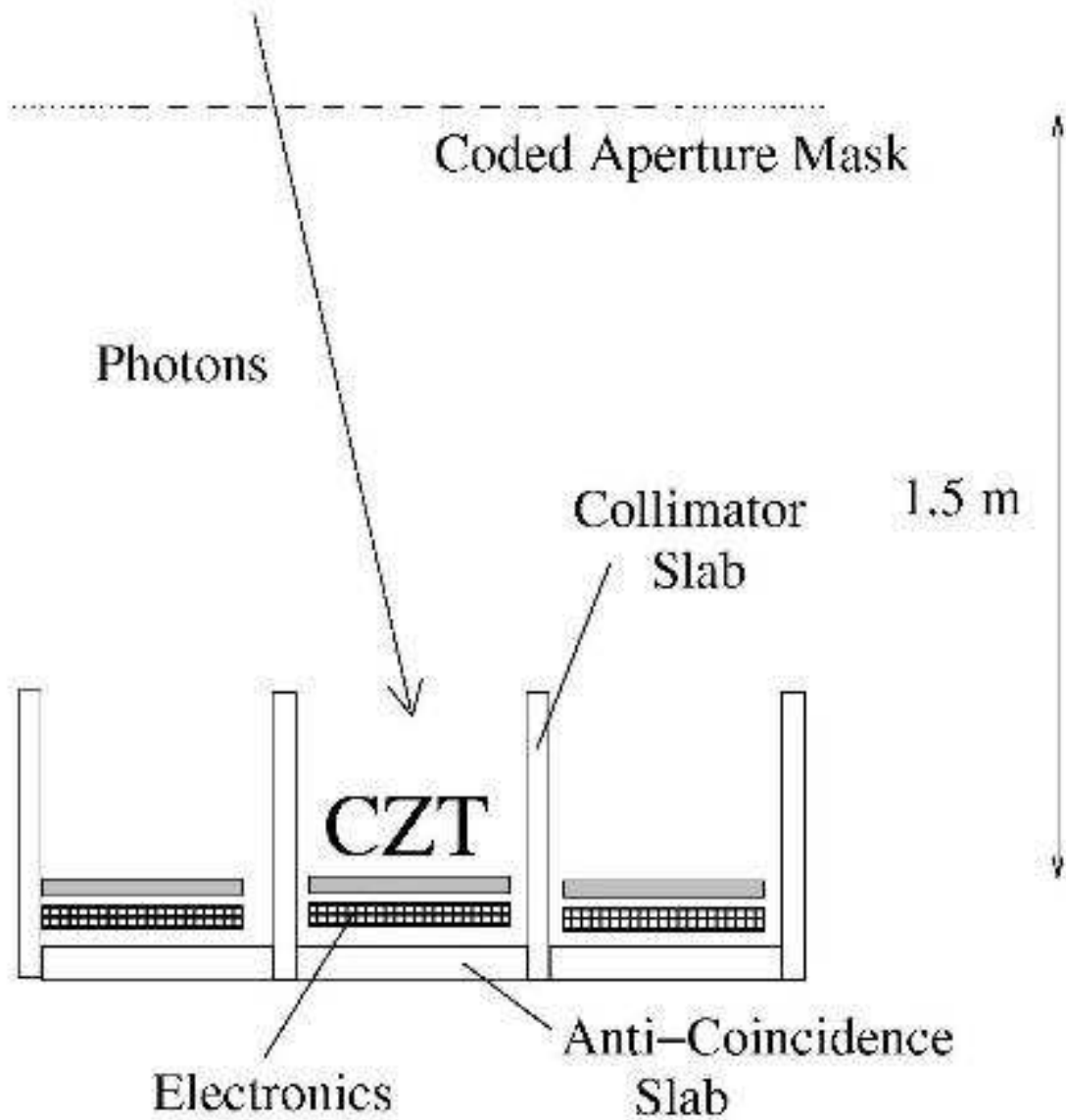

**Fig. 2.** Side view of an *EXIST*-type telescope showing three sub-telescopes with coded mask, CZT detectors, and shield/collimator. The shadow pattern produced by the coded mask is seen by the CZT detectors. This pattern can be deconvoluted to reconstruct an X-ray image. The active shield/collimator defines the field-of-view for the CZT detectors and reduces the instrumental background .

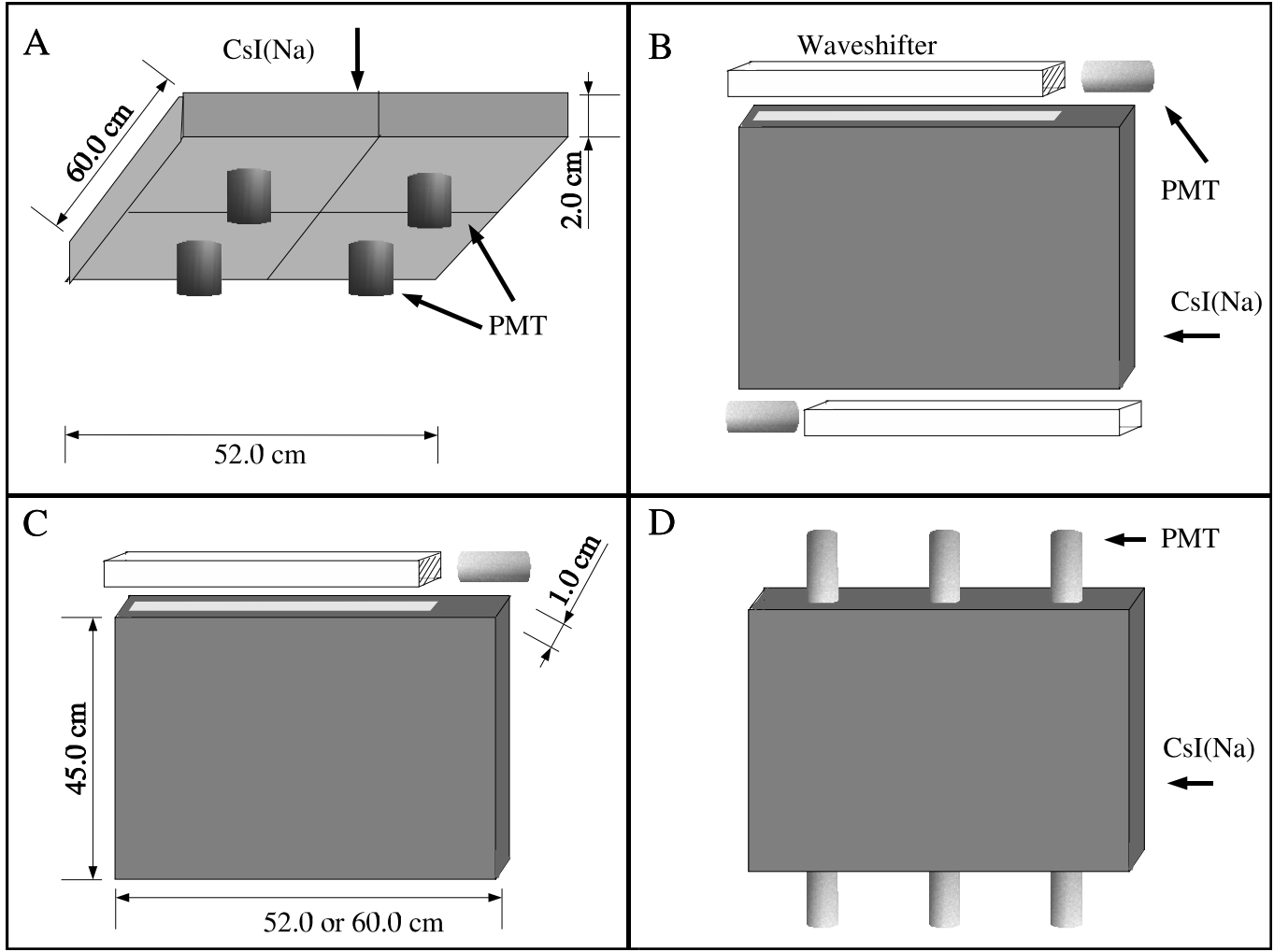

**Fig. 3.** Panel A shows the configuration to read out the four segments of a bottom slab with four PMTs. Panels B-D show three different configurations to read out a side slab. We find that configurations B and D give very different performance, though configuration D uses 3 times the power of configuration B (See Figure 4).

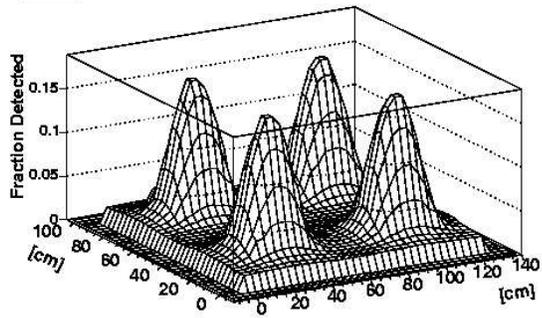
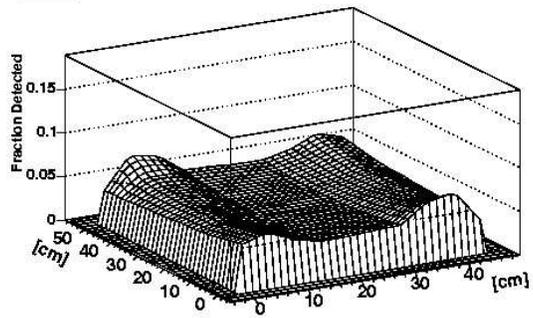
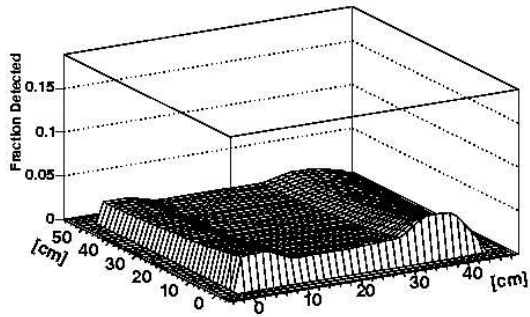
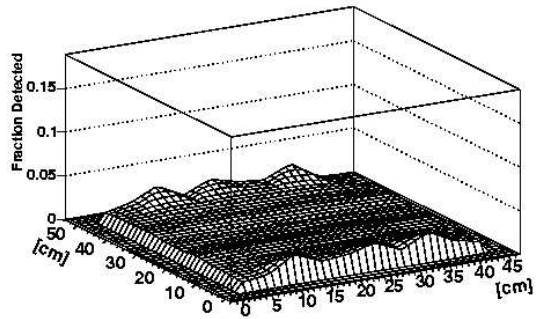

**Fig. 4.** Fraction of scintillation photons reaching the photodetectors as function of the location of photon generation for configurations A-D of Figure 3.

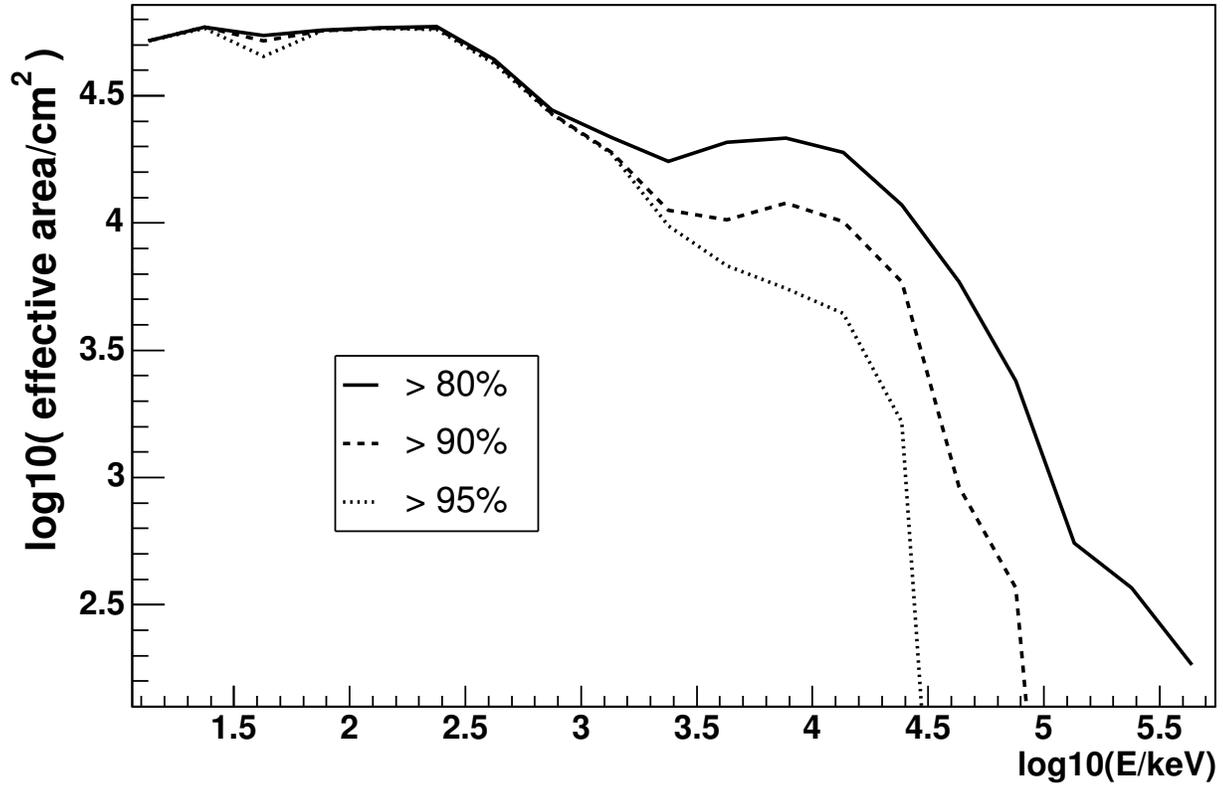

**Fig. 5.** Effective area of the shield/collimator assembly as a function of the energy of the incident photon. A photon counts as "detected" if it deposits more than 80% (dotted line), 90% (dashed line), or 95% (solid line) of incident energy in any one slab.

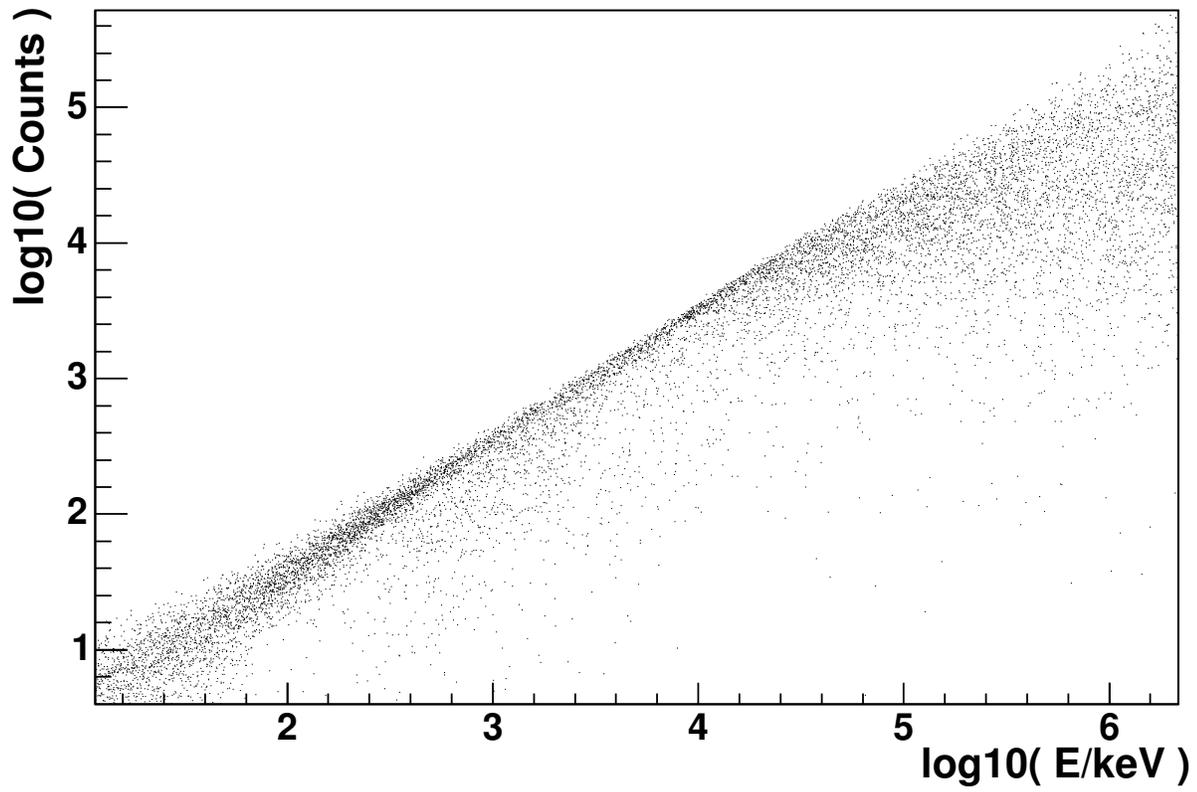

**Fig. 6.** PMT photoelectrons as a function of incident photon energy. The energy resolution is limited by photon statistics at low energies and leakage at higher energies.

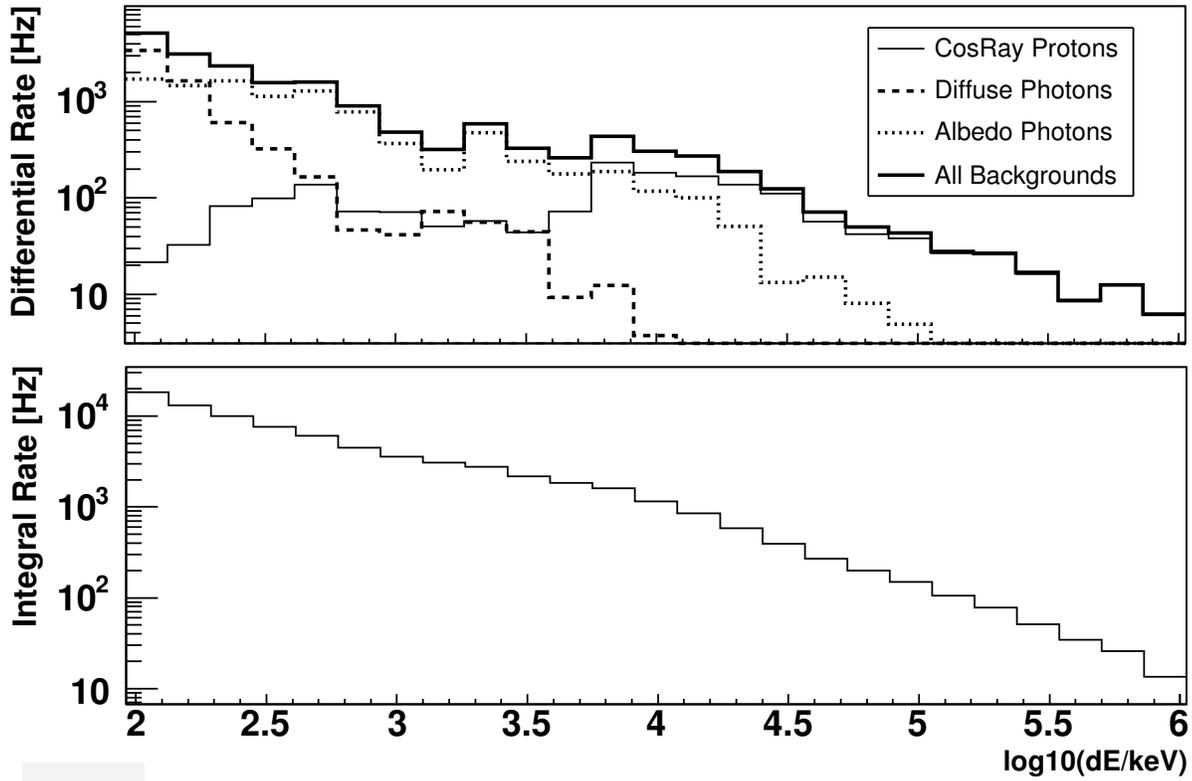

**Fig. 7.** Simulated background rates for an *EXIST* sub-telescope shield/collimator. A sub-telescope shield consists of four side slabs and a rear shield comprised of four bottom slabs. The top panel shows the differential background rate as a function of energy deposited in the shield. The bottom panel gives the integral background count rate as a function of the shield low-energy threshold.

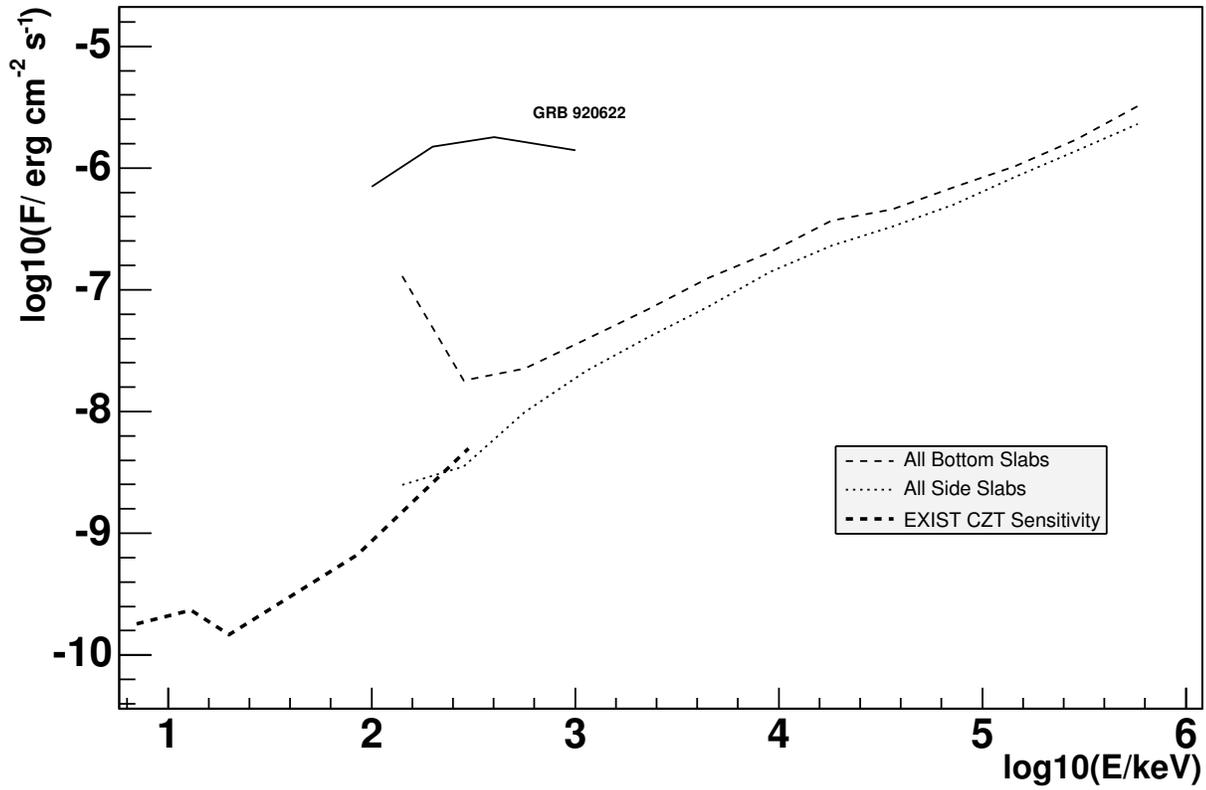

**Fig. 8.** GRB flux sensitivity of the *EXIST* shield/collimator assembly for factor-of-2 wide energy intervals (Dashed line: all bottom slabs; Dotted line: all side slabs). For comparison, the flux sensitivity of the CZT detectors (bold dashed line), and the spectrum of the strong GRB 920622 (solid line, from Greiner et al. (2003)) are also shown.

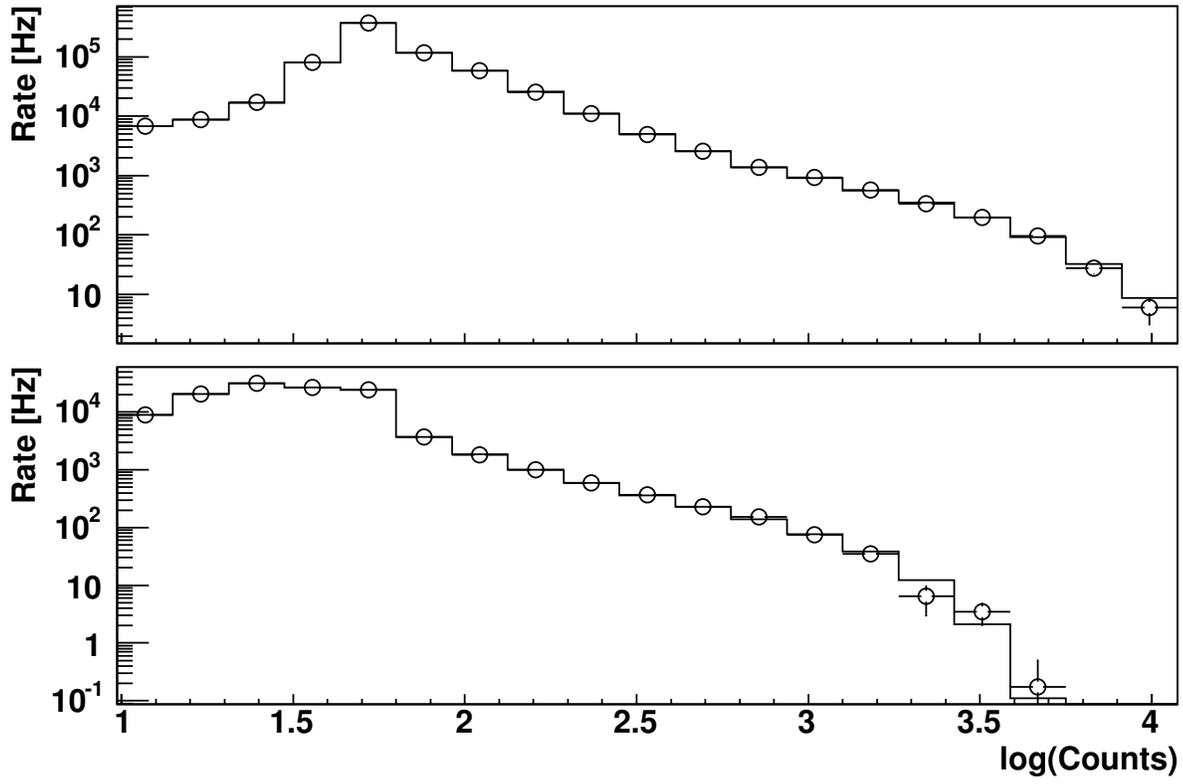

**Fig. 9.** Cumulative PMT pulse height histograms are shown for all side slabs (upper) and all bottom slabs (lower) for a GRB simulated with the parameters $E_P = 450$ keV and $N_P = 10^{-7}$ ergs cm$^{-2}$ s$^{-1}$. The points with error bars show the detected GRB counts and the histograms show the best-fit model.

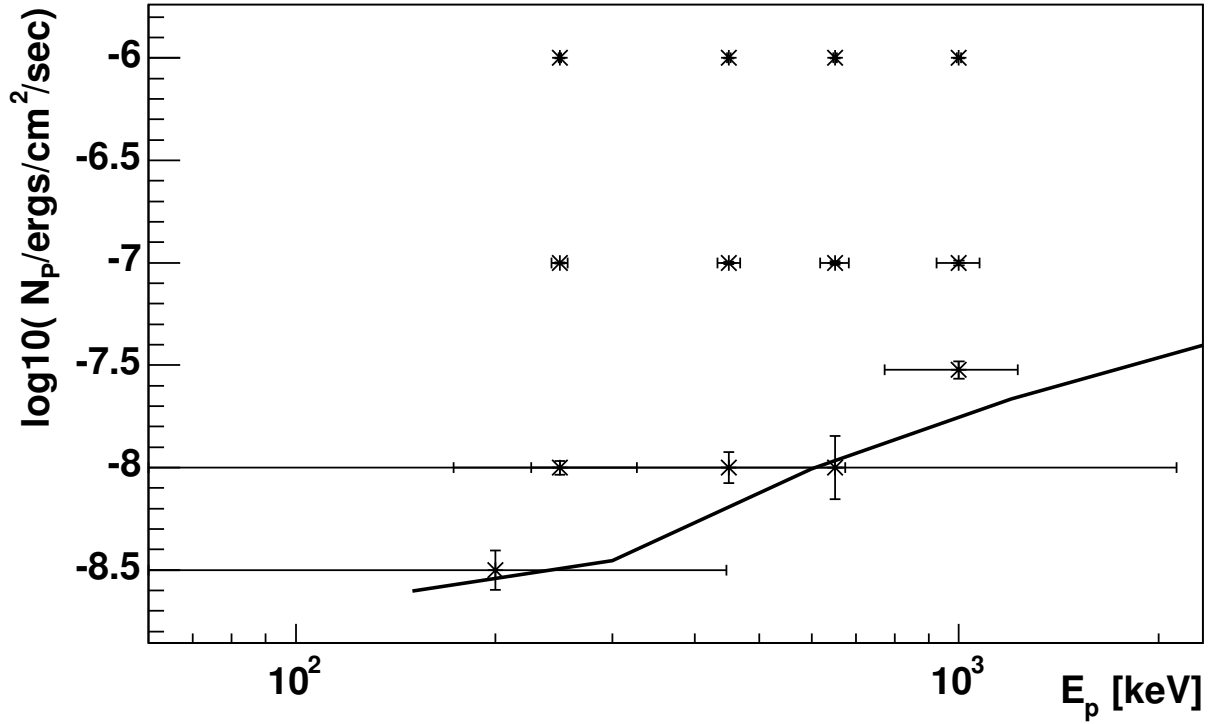

**Fig. 10.** Fitting results for 12 different GRB spectra. The stars show the 12 parameter combinations $E_P$ and $N_P$ for which we simulated artificial GRBs. The error bars show the accuracy with which the parameters can be reconstructed from the simulated data (90% of the reconstructed $E_P$ and $N_P$ were found within the error bars). The solid line (lower right) shows the sensitivity of the side slabs from Fig. 8.

| Properties of CsI(Na) | All Slabs |
|---|---|
| Density (g/cm$^2$) | 4.61 |
| Peak Scintillation Wavelength (nm) | 420 |
| Photons/MeV | 41 000 |
| Refractive Index | 1.78 |

| Slab Dimensions | [cm$^3$] |
|---|---|
| Bottom (1/4 section) | 25×29×2 |
| Bottom (combined) | 50×58×2 |
| Small Side | 45× 50×1 |
| Big Side | 45×58×1 |

| PMT properties | (side slabs) |
|---|---|
| Diameter | 1.25 cm |
| Quantum Efficiency | 15% |

| PMT properties | (bottom slabs) |
|---|---|
| Diameter | 7.5 cm |
| Quantum Efficiency | 15% |

| Waveshifter Bar Properties | Side Slabs |
|---|---|
| Larger Side Slab | $1.25 \times 1.25 \times 50.05$ cm$^3$ |
| Smaller Side Slab | $1.25 \times 1.25 \times 47.0$ cm$^3$ |
| Wavelength Shifted Value | 500 nm |
| Wavelength Shifting Mean Free Path | 2.0 mm |
| Refractive Index | 1.6 |

**Table 1.** Properties of CsI(Na) Slabs, waveshifters, and PMTs

| Simulated $E_P$ | Simulated $\log10(N_P)$ | Avg. Fit $E_P \pm 90\%$ range | Avg. Fit $\log10(N_P) \pm \log10(90\%)$ range |
|---|---|---|---|
| 200 | -8.4 | $195^{+339}_{-146}$ | $-8.35^{+.08}_{-.07}$ |
| 250 | -6 | $250^{+1}_{-1}$ | $-6.00^{+.001}_{-.001}$ |
| 250 | -7 | $250^{+7}_{-7}$ | $-7.00^{+.004}_{-.003}$ |
| 250 | -8 | $237^{+47}_{-106}$ | $-8.00^{+.03}_{-.03}$ |
| 450 | -6 | $450^{+3}_{-3}$ | $-6.0^{+.001}_{-.001}$ |
| 450 | -7 | $447^{+17}_{-19}$ | $-7.00^{+.005}_{-.005}$ |
| 450 | -8 | $398^{+234}_{-214}$ | $-8.02^{+.08}_{-.06}$ |
| 650 | -6 | $650^{+6}_{-6}$ | $-6.00^{+.001}_{-.001}$ |
| 650 | -7 | $648^{+33}_{-33}$ | $-7.00^{+.008}_{-.01}$ |
| 650 | -8 | $575^{+2610}_{-346}$ | $-8.02^{+.12}_{-.09}$ |
| 1000 | -6 | $1000^{+8}_{-8}$ | $-6.00^{+.001}_{-.001}$ |
| 1000 | -7 | $998^{+65}_{-85}$ | $-7.00^{+.014}_{-.008}$ |
| 1000 | -7.5 | $942^{+230}_{-224}$ | $-7.52^{+.04}_{-.05}$ |

**Table 2.** Spectral Fits Model Parameters and Results - For a given $E_P$ and $N_P$, the average fit for 100 simulated bursts. The errors give the 90% range. $E_P$ are given in keV and $N_P$ are given in ergs cm$^{-2}$s$^{-1}$.

| Model | Parameter | Simulated Value | Avg. Fit |
|-------|-----------|-----------------|----------|
| Band | $\alpha$ | $-1.0$ | $-0.99^{+.052}_{-.049}$ |
| Band | $\beta$ | $-2.5$ | $-2.47^{+.052}_{-.092}$ |
| Thermal | kT | 100 keV | $100.3^{+2.1}_{-1.6}$ keV |
| Thermal | log10(Normalization) | $-7.0$ ergs cm$^{-2}$ s$^{-1}$ | $-7.0^{+.01}_{-.01}$ ergs cm$^{-2}$ s$^{-1}$ |

**Table 3.** Spectral Fits Model Parameters and Results II - For a given $E_P$, $N_P$, $\alpha$, and $\beta$, the average fit for 100 simulated bursts' spectral indices is shown. Also, For a given kT and amplitude, the average fit for 100 simulated bursts is shown. The errors give the 90% range. For the Band model fits, $E_P = 450$ keV and $N_P = 10^{-7}$ergs cm$^{-2}$s$^{-1}$.